\documentclass[useAMS,usenatbib]{mn2e}

\usepackage{psfig}
\usepackage{graphicx}
\usepackage{times}
\usepackage{natbib}

\newif\ifAMStwofonts
\AMStwofontstrue

\newcommand{\be}{\begin{equation}}
\newcommand{\ee}{\end{equation}}
\newcommand{\ba}{\begin{eqnarray}}
\newcommand{\ea}{\end{eqnarray}}
\newcommand{\brr}{\begin{array}}
\newcommand{\err}{\end{array}}
\newcommand{\bc}{\begin{center}}
\newcommand{\ec}{\end{center}}
\newcommand{\hm}{\,h^{-1}{\rm Mpc}}

\newcommand{\msun}{\,h^{-1}M_\odot}

\newcommand{\vel}{\,{\rm km\,s^{-1}}}
\newcommand{\hub}{\,{\rm km\,s^{-1}Mpc^{-1}}}

\newcommand{\mincir}{\raise
  -2.truept\hbox{\rlap{\hbox{$\sim$}}\raise5.truept \hbox{$<$}\ }}
\newcommand{\magcir}{\raise
  -2.truept\hbox{\rlap{\hbox{$\sim$}}\raise5.truept \hbox{$>$}\ }}
\newcommand{\siml}{\raise
  -2.truept\hbox{\rlap{\hbox{$\sim$}}\raise5.truept \hbox{$<$}\ }}
\newcommand{\simg}{\raise
  -2.truept\hbox{\rlap{\hbox{$\sim$}}\raise5.truept \hbox{$>$}\ }}

\title[Angular diameter distance estimates from the SZ effect]
{Angular diameter distance estimates from the Sunyaev--Zeldovich
effect in hydrodynamical cluster simulations} \author[Ameglio et al.]
{S. Ameglio$^{1,2,3}$, S. Borgani$^{1,2,3}$, A. Diaferio$^4$ \&
K. Dolag$^5$ \\~\\ $^1$ Dipartimento di Astronomia dell'Universit\`a
di Trieste, via Tiepolo 11, I-34131 Trieste, Italy
(ameglio,borgani@ts.astro.it) \\ $^2$ INFN -- National Institute for
Nuclear Physics, Trieste, Italy\\ $^3$ INAF -- National Institute for
Astrophysics, Trieste, Italy\\ $^4$ Dipartimento di Fisica Generale
``Amedeo Avogadro'', Universit\`a degli Studi di Torino, Via P. Giuria
1, I-10125, Torino, Italy\\ (diaferio@ph.unito.it)\\ $^5$
Max-Planck-Institut f\"ur Astrophysik, Karl-Schwarzschild Strasse 1,
Garching bei M\"unchen, Germany (kdolag@mpa-garching.mpg.de) }

\begin{document}

\date{Accepted ???. Received ???; in original form ???}

\maketitle

\begin{abstract}
The angular--diameter distance $D_A$ of a galaxy cluster can be
measured by combining its X--ray emission with the cosmic microwave
background fluctuation due to the Sunyaev--Zeldovich effect.  The
application of this distance indicator usually assumes that the
cluster is spherically symmetric, the gas is distributed according to
the isothermal $\beta$--model, and the X--ray temperature is an
unbiased measure of the electron temperature.  We test these
assumptions with galaxy clusters extracted from an extended set of
cosmological N-body/hydrodynamical simulations of a $\Lambda$CDM
concordance cosmology, which include the effect of radiative cooling,
star formation and energy feedback from supernovae.  We find that, due
to the temperature gradients which are present in the central regions
of simulated clusters, the assumption of isothermal gas leads to a
significant underestimate of $D_A$.  This bias is efficiently
corrected by using the polytropic version of the $\beta$--model to
account for the presence of temperature gradients. In this case, once
irregular clusters are removed, the correct value of $D_A$ is
recovered with a $\sim 5$ per cent accuracy on average, with a $\sim
20$ per cent intrinsic scatter due to cluster asphericity. This result
is valid when using either the electron temperature or a
spectroscopic--like temperature.  When using instead the
emission--weighted definition for the temperature of the simulated
clusters, $D_A$ is biased low by $\sim 20$ per cent. We discuss the
implications of our results for an accurate determination of the
Hubble constant $H_0$ and of the density parameter $\Omega_m$.  
We find that, at least in the case of ideal (i.e. noiseless) X-ray
and SZ observations extended out to $r_{500}$, $H_0$ can be potentially
recovered with exquisite precision, while the resulting estimate of
$\Omega_m$, which is unbiased, has typical errors
$\Delta\Omega_m\simeq 0.05$.
\end{abstract}

\begin{keywords}
large-scale structure of Universe -- galaxies: clusters: general -- cosmology:
miscellaneous -- methods: numerical
\end{keywords}

\section{Introduction}
The Sunyaev--Zeldovich (SZ) effect \citep{1972CoASP...4..173S} is the
distortion of the Cosmic Microwave Background (CMB) spectrum due to
the scattering of the CMB photons off a population of electrons. At
radio frequencies, the typical size of this distortion for a thermal
distribution of electrons with temperature of about 10 keV is at the
level of $10^{-4}$. This effect has now been detected for a fairly
large number of clusters of galaxies \citep[e.g.][ for
reviews]{1995ARA&A..33..541R,1999PhR...310...97B,2002ARA&A..40..643C}. For
almost three decades it has been recognized that the combination of
X--ray and SZ observations of galaxy clusters provides a direct
measurement of the cosmic distance scale, under the assumption of
spherical symmetry for the intra--cluster gas distribution
\citep[e.g.][]{1978obco.meet....1G,1978ApJ...226L.103S,1979A&A....75..322C,1979MNRAS.187..847B}. The
method is based on the different dependence on the electron number
density, $n_e$, of the X--ray emissivity ($\propto n_e^2T_e^{1/2}$ for
thermal bremsstrahlung; here $T_e$ is the electron temperature) and of
the SZ signal ($\propto n_e T_e$).

Due to the crucial role played by the assumption of spherical
symmetry, a great deal of efforts have been spent either to select
individual clusters having very relaxed and regular morphology
\citep[e.g.][]{1997ApJ...480..449H,1998ApJ...501....1H,2002MNRAS.333..318G,2004ApJ...614...56B},
or to build suitable cluster samples over which averaging out the
uncertainties due to intrinsic cluster ellipticity
\citep[e.g.][]{2001ApJ...555L..11M,2002ApJ...581...53R,2004ApJ...615...63U,2005MNRAS.357..518J}. These
analyses have provided estimates of the Hubble constant, $H_0$, which
are generally consistent with those obtained from the Cepheid distance
scale \citep[e.g.][]{2001ApJ...553...47F} or inferred from the
spectrum of the CMB anisotropies
\citep[e.g.][]{SP03.1}, although with fairly large
uncertainties. Although the dominant source of uncertainty is probably
represented by the contamination of the SZ signal by the CMB and
point--sources \citep[e.g.][]{2004ApJ...615...63U}, significant errors
are also associated to cluster asphericity, clumpy gas distribution
and incorrect modeling of the thermal structure of the intra--cluster
medium (ICM).

So far, the limited number of high--redshift clusters with both SZ and
X--ray observations, with their relatively large uncertainties, made
the calibration of the cosmic distance scale mostly sensitive to the
value of $H_0$, while no significant constraints have been placed on
the values of the matter density parameter $\Omega_m$ and the
cosmological constant.  In the coming years, ongoing X--ray
\citep[e.g.,][]{2005ApJ...623L..85M}, optical
\citep[e.g.,][]{2005ApJS..157....1G}, and planned or just started SZ
surveys\footnote{See, for example, the dedicated interferometer
arrays:
\begin{itemize}{\scriptsize
\item AMI: {\tt http://www.mrao.cam.ac.uk/telescopes/ami/index.html}
\item AMiBA: {\tt http://www.asiaa.sinica.edu.tw/amiba}
\item SZA: {\tt http://astro.uchicago.edu/sze}
}
\end{itemize}
or the bolometers:
\begin{itemize}{\scriptsize
\item ACBAR: {\tt http://cosmology.berkeley.edu/group/swlh/acbar/}
\item ACT: {\tt http://www.hep.upenn.edu/$\sim$angelica/act/act.html}
\item APEX: {\tt http://bolo.berkeley.edu/apexsz}
\item Olimpo: {\tt http://oberon.roma1.infn.it/}
\item Planck: {\tt http://astro.estec.esa.nl/Planck/}
\item SPT: {\tt http://astro.uchicago.edu/spt/}
}
\end{itemize}
}  promise to largely increase the number of
observed clusters out to $z\sim 1.5$. This may
well open the possibility to use SZ/X--ray cluster observations to
place constraints on the Dark Matter and Dark Energy content of the
universe \citep{2002ApJ...570....1M}. This highlights the paramount
importance of having observational uncertainties under control.

In this respect, numerical hydrodynamical simulations of galaxy clusters
may offer an important test--bed where to quantify observational
biases and keep the corresponding uncertainties under control. For
instance, eliminating $n_e$ from the SZ and X--ray signal leaves a
sensitive dependence of the angular--diameter distance, $D_A$, on the
electron temperature (see \S 2). On the other hand, temperature
measurements of the ICM have been so far entirely based on fitting the
X--ray spectrum to a suitable plasma model. How close is the resulting
spectral temperature to the electron temperature depends on the
complexity of the thermal structure of the ICM
\citep[e.g.,][]{2004MNRAS.354...10M}. Hydrodynamical simulations of
clusters offer a natural way to quantify the bias introduced by
replacing the electron temperature with the X--ray
temperature. Furthermore, the standard assumption in the SZ/X--ray
calibration of the cosmic distance scale is that of the isothermal
ICM, while X--ray observations of clusters clearly show the presence
of significant temperature gradients
\citep[e.g.][]{1998ApJ...503...77M,2002ApJ...567..163D,2005ApJ...628..655V}.
To overcome this problem, several authors
estimate the bias introduced by the isothermal approximation, finding
that the distance can be biased by $ \mincir 20$ per cent
\citep[e.g.][]{1994ApJ...420...33B,2004ApJ...615...63U,1997ApJ...480..449H}.
Simulations of galaxy clusters naturally produce temperature gradients
that, at least at large radii, are close to the observed ones
\citep[e.g.,][]{2002ApJ...579..571L,2004MNRAS.348.1078B,2004MNRAS.355.1091K}. 
Therefore,
simulations can be used to quantify the bias introduced by the
assumption of isothermal gas. Finally, using a representative set of
galaxy clusters in a cosmological framework allows one to trace the
distribution of ellipticity and, therefore, to calibrate the
corresponding scatter in the measurement of the distance
scale. Nowadays, cosmological hydrodynamical codes have reached a high
enough efficiency, in terms of both achievable resolution and
description of the gas physics, to provide a realistic description of
the processes of formation and evolution of galaxy clusters
\citep[e.g.][]{2004MNRAS.348.1078B,2005ApJ...625..588K}. For instance,
\citet{2004ApJ...611L..73K} found that halos in hydrodynamical
simulations including cooling are significantly more spherical than in
non--radiative simulations. Since the assumption of sphericity is at
the basis of the X--ray/SZ method to estimate $D_A$, this highlights
the relevance of properly modeling the physics of the ICM for a
precise calibration of the cosmic distance scale.

The purpose of this paper is to understand the impact of the above
discussed systematics on the calibrations of the cosmic distance scale
from the combination of SZ and X--ray observations, by analyzing an
extended set of hydrodynamical simulations of galaxy clusters. These
simulations have been performed using the TREE+SPH {\small GADGET--2}
code \citep{SP01.1,2005astro.ph..5010S}, for a concordance
$\Lambda$CDM model, and include the processes of radiative cooling,
star formation and supernova feedback. The set of simulated clusters
contains more than 100 objects having virial masses in the range
$(2-20)\times 10^{14}h^{-1} M_{\odot}$.

The plan of the paper is as follows. In Section 2 we review the method
to estimate the angular--diameter distance from X--ray and SZ cluster
observations in the case of a polytropic equation
of state for the ICM, and discuss the different definitions of
temperature that are used in the analysis of the simulations. In Section 3
we describe the set of simulated clusters and the procedure to
generate X--ray and SZ maps. We present our results in Section 4,
where we show the results on the accuracy of the measurement of
$D_A$. We discuss and summarize our main results in Section 5.

\section{$D_A$ from combined X-ray and SZ observations}
The combination of the ICM X-ray emission with the SZ flux decrement
provides a direct measure of the angular-diameter distance, $D_A$,
of galaxy clusters
\citep{1978ApJ...226L.103S,1979A&A....75..322C,1979MNRAS.187..847B}. The
method takes advantage of the different dependences of these two
quantities on the electron number density, $n_e$ (the former is
$\propto n_e^2$ while the latter is $\propto n_e$).
When combined, these two quantities provide the physical dimension of the 
cluster along the line-of-sight and, from the cluster angular size, $D_A$,
if the cluster is spherically symmetric.

The Comptonization parameter $y$, as measurable from observations of
the SZ effect, is 
\be y=\int n_e
\frac{k_B T_e}{m_e c^2} \sigma_T d\ell 
\label{eq:y}
\ee 
where $k_B$ is the Boltzmann constant, $\sigma_T$ is the Thompson
cross section, $m_e$ is the mass of the electron, $c$ is the speed of
light and the integration is along the line of sight. By
definition, it provides a redshift--independent measure of the total
thermal content of the cluster.

The X-ray surface brightness $S_X$ is 
\be
S_X=\frac{1}{4\pi(1+z)^4}\int n_en_H\Lambda_{eH}(T)d\ell 
\label{eq:sx}
\ee 
where $n_H$ is the hydrogen number density of the ICM,
$\Lambda_{eH}(T)$ is the cooling function (normalized to $n_e
n_H$). To compute the X--ray emissivity of the simulated clusters, 
we assume the cooling function taken from a
Raymond-Smith code \citep{1977ApJS...35..419R} for a gas of primordial
composition ($X_H=0.76$ and $X_{He}=0.24$) for a fully ionized ICM.  
We compute the X--ray emissivity 
in the [0.5--2] keV energy band, which is used in
several combined SZ/X--ray analyses relying on the ROSAT--PSPC data
for the X--ray imaging part
\citep[e.g.][]{2002ApJ...581...53R,2005MNRAS.357..518J}. Using bands
extending to higher energies, as appropriate for Chandra and
XMM--Newton observations, would produce no change in the final results.

\subsection{The polytropic $\beta$--model}
A common procedure adopted to extract $D_A$ from the combination of
eqs. (\ref{eq:y}) and (\ref{eq:sx}) is based on modeling the electron
density profile with the $\beta$-model,
\be\label{eq:ne}
n_e(r)=n_{e0}\left[1+\left(\frac{r}{r_c}\right)^2\right]^{-3\beta/2}
\ee
\citep{1976A&A....49..137C}, where $n_{e0}$ is the electron number
density in the cluster center, $r$ is the distance from the cluster
center, $r_c$ is the core radius and $\beta$ is the power--law index.

As for the temperature structure of the ICM, a number of analyses of
X--ray data independently show that galaxy clusters are far from being
isothermal. Significant negative gradients characterize the
temperature profiles of galaxy clusters, at least on scales $R\magcir
0.2 R_{200}$
\citep[e.g.][]{1998ApJ...503...77M,2002ApJ...567..163D,2002A&A...394..375P,2005A&A...433..101P,2005ApJ...628..655V},
with positive gradients associated only to the innermost cooling
regions \citep[e.g.][]{2001MNRAS.328L..37A}. The dynamic range covered
by the SZ signal extends on scales which are relatively larger than
those sampled by the X--ray emission. For this reason, one may expect
that a systematic effect is introduced by assuming the ICM to be
isothermal when combining X--ray and SZ observations. Since the SZ
signal has a stronger dependence on the ICM temperature than the
X--ray one, the effect of assuming an isothermal ICM, in a regime
where the temperature is decreasing, may lead to predict
$y(\theta)$--profiles which are shallower than the intrinsic ones.

In order to account for the presence of temperature gradients, we
introduce a polytropic equation of state, $p\propto \rho^\gamma$,
which relates the gas pressure $p$ to the density $\rho$, where $\gamma$ is
the polytropic index ($\gamma=1$ for isothermal gas). The
three-dimensional temperature profile is thus 
\be
T_e(r)=T_{e0}\left(n_e\over n_{e0}\right)^{\gamma-1}=T_{e0}
\left[1+\left(\frac{r}{r_c}\right)^2\right]^{-3\beta(\gamma-1)/2}, 
\label{eq:tpol}
\ee
where $T_{e0}$ is the temperature at the cluster center. Using the
above expression for the temperature profile in the definition of
the Comptonization parameter of eq.(\ref{eq:y}) gives
\be
y(\theta)=y_0\left[1+\left(\frac{\theta}{\theta_c}\right)^2\right]^{(1-3\beta\gamma)/2}\,,
\label{eq:ypol}
\ee
where the Comptonization parameter at the cluster center is
\be
y_0=D_An_{e0}\theta_c \sigma_T{k_BT_{e0}\over m_ec^2}
\sqrt{\pi}{\Gamma(3\beta\gamma/2-1/2) \over \Gamma(3\beta\gamma/2)}\,. 
\label{eq:y0pol}
\ee

Similarly, we obtain the X--ray surface brightness profile
\be
S_X(\theta)=S_{X0}\left[1+\left(\frac{\theta}{\theta_c}\right)^2\right]^
{\left\{1-6\beta [(\gamma +3)/4]\right\}/2}\,,
\label{eq:sxpol}
\ee
where the central surface brightness is
\be S_{X0}=D_An_{e0}^2 \theta_c{1\over 4\sqrt\pi}{\mu_e \over
\mu_H}\Lambda_{eH}(T_{e0}){\Gamma(3\beta[(\gamma+3)/4]-1/2)\over
\Gamma(3\beta[(\gamma+3)/4]}\,.
\label{eq:sx0pol}
\ee
In deriving the above equation, the dependence of the cooling
function on $T_e$ is assumed to be a power law, $T^\alpha$, with index
$\alpha = 0.5$. This is valid in the case of pure bremsstrahlung
emission and represents a good approximation in the case of bolometric
emissivity. However, our emissivity maps are build in the [0.5-2] keV
band. In this energy range the cooling function is significantly
flatter, due to the contribution of metal emission lines, which is 
relevant in the case of relatively cool systems ($T_e< 2$ keV)
\citep[e.g.][]{2000MNRAS.311..313E}. In order to test the effect of
approximating the cooling function with a bremsstrahlung shape, we
repeated our analysis also in the bolometric band and found variations
in the final distance estimates by $\mincir 10\%$.

Finally, by eliminating $n_{e0}$ from eqs.(\ref{eq:y0pol}) and
(\ref{eq:sx0pol}), we obtain the angular--diameter distance
\begin{eqnarray}
D_A&=&\frac{y_0^2}{S_{X0}}\left[\frac{m_ec^2}{k_BT_{e0}}\right]^2\frac{\Lambda_{eH}(T_{e0})\mu_e/\mu_H}{4\pi^{3/2}\sigma_T^2(1+z)^4}\frac{1}{\theta_c}\nonumber \\
&&\times \left[\frac{\Gamma(3\beta\gamma/2)}{\Gamma(3\beta/2-1/2)}\right]^2\frac{\Gamma(3\beta[(\gamma+3)/4]-1/2)}{\Gamma(3\beta[(\gamma+3)/4])}\,.
\label{eq:dapol}
\end{eqnarray}
For $\gamma=1$, the above expression reduces to that usually adopted
in observational analyses based on combining X--ray and SZ cluster
observations
\citep[e.g.][]{2002ApJ...581...53R,2004ApJ...615...63U,2004ApJ...614...56B},
which relies on the assumption of isothermal gas.

It is worth reminding here that, while simulations are rather
successful in reproducing the negative temperature gradient in the
outer cluster regions
\citep[e.g.][]{1996ApJ...469..494E,1998ApJ...503..569E,2002ApJ...579..571L,2004MNRAS.351..237R},
they generally produce too steep profiles in the central cluster
regions, especially when radiative cooling is included
\citep[e.g.][]{1993ApJ...412..455K,2003MNRAS.342.1025T,
2003MNRAS.339.1117V,2004MNRAS.348.1078B}. Since observed clusters are
characterized by a core region which is closer to isothermality than the
simulated ones, we expect that the effect of using a polytropic
temperature profile when analyzing simulated clusters is larger than the actual
effect taking place in real clusters.

\subsection{Definitions of temperature}
In the observational determinations of $D_A$ from X--ray and SZ
observations with eq.(\ref{eq:dapol}) one relies on the X--ray
temperature obtained from the spectral fitting. Such a spectral
temperature is generally different from both the actual electron
temperature, which appears in the expression of $D_A$, and the
emission--weighted temperature, which is often used as a proxy to the
spectral temperature in the analysis of hydrodynamical simulations of
clusters \citep[e.g.][]{1996ApJ...469..494E}.

If $n_{e,i}$ and $T_i$ are defined as the electron number density and
the temperature carried by the $i$--th simulation gas particle, then
the electron temperature is defined by 
\be
T_e=\sum_i n_{e,i}T_i/\sum_i n_{e,i}\,,
\label{eq:tel}
\ee
which coincides with the mass--weighted temperature in the limit of a
fully ionized plasma of uniform metallicity. Analogously, 
the emission--weighted temperature is 
\be
T_{ew}={\sum_i  \Lambda (T_i) n_{e,i}^2 T_i \over \sum_i \Lambda(T_i)
  n_{e,i}^2}\,,
\label{eq:tew}
\ee
where the cooling function $\Lambda(T)$ can be computed over an
energy band, comparable to that where the
X--ray spectrum is fitted in observational data analyses. In the
following, we compute the emissivity in the [0.5--7] keV band.

However, \cite{2001ApJ...546..100M} were the first to show that the
emission--weighted temperature does not necessarily represent an
accurate approximation to the spectroscopic temperature.
\cite{2004MNRAS.354...10M} have further motivated and quantified this
difference, connecting it to a thermally complex structure of the
ICM. These authors suggested an approximate expression for the
spectroscopic temperature, the spectroscopic--like temperature:
\be
T_{sl}=\frac{\sum_i n_{e,i}^2 T_i^{a+1/2}}{\sum_i n_{e,i}^2 T_i^{a+3/2}}\,,
\label{eq:tsl}
\ee
where $a$ is a fitting parameter. \cite{2004MNRAS.354...10M} have
shown that eq.(\ref{eq:tsl}) with $a=0.75$ closely reproduces the
spectroscopic temperature of clusters at least as hot as 3 keV,
with a few per cent accuracy, after excluding all the gas particles
colder than 0.5 keV from the sums in eq.(\ref{eq:tsl}).  More
recently, \citep{2005astro.ph..4098V} has generalized the above
expression for $T_{sl}$ to include the cases of lower temperatures and
arbitrary ICM metallicity.

In the following, besides using the electron temperature, we also
perform our analysis by relying on the temperature proxies of
eqs.(\ref{eq:tew}) and (\ref{eq:tsl}). Therefore, comparing the
results based on the electron temperature and on the
spectroscopic--like temperature provides a check of the bias
introduced by using the X--ray temperature in the estimate of $D_A$, a
bias possibly present also in the analysis of real data. Furthermore,
the comparison between emission--weighted and spectroscopic--like
temperature provides a hint on the bias introduced in the simulation
analysis when using an inaccurate proxy to the X--ray
temperature. It is worth reminding here that, due to the finite
time for electron--ion thermalization, the corresponding electron and
ion temperature may differ, for instance as a consequence of
continuous shocks \citep[e.g.,][]{2005ApJ...618L..91Y}. A sizable difference
among these two temperatures may induce a bias in the estimate of the
distance scale.

Except for using different definitions of temperature, we do not
investigate the effect of a realistic observational setup for the
detection of both the SZ and X--ray signals. Besides the statistical
errors associated to time--limited exposures, we also neglect the
effects of systematics (e.g., instrumental noise, foreground and
background contribution from contaminants, etc.). A detailed analysis
of the contaminations in the SZ signal has been provided by
\cite{2004ApJ...612...96K} and by \cite{2004astro.ph..2571A}. A
comprehensive description of the instrumental effects on the recovery
of X--ray observables, calibrated on hydrodynamical simulations, has
been provided by \cite{2004MNRAS.351..505G} \citep[see also][] 
{2006astro.ph..2434R}. In this sense, our analysis will be
based on ideal maps, which are free of any noise. We defer to a future
analysis the inclusion of the errors associated to realistic X--ray
and SZ observational setups.

\section{The simulated clusters}

The sample of simulated galaxy clusters used in this paper has been
extracted from the large-scale cosmological hydro-N-body simulation of
a ``concordance'' $\Lambda$CDM model with $\Omega_m=0.3$ for the
matter density parameter at present time, $\Omega_\Lambda=0.7$ for the
cosmological constant term, $\Omega_{\rm b}=0.019\,h^{-2}$ for the
baryons density parameter, $h=0.7$ for the Hubble constant in units of
100 km s$^{-1}$Mpc$^{-1}$ and $\sigma_8=0.8$ for the r.m.s. density
perturbation within a top--hat sphere having comoving radius of
$8\hm$. We refer to \cite{2004MNRAS.348.1078B} (B04 hereafter) for a
detailed presentation of this simulation, while we give here only a
short description.

The run, performed with
the massively parallel Tree+SPH code {\small GADGET-2}
\citep{SP01.1,2005astro.ph..5010S}, follows the evolution of $480^3$
dark matter particles and an equal number of gas particles in a
periodic cube of size $192 h^{-1}$ Mpc. The mass of the gas particles
is $m_{\rm gas}=6.9 \times 10^8 h^{-1} M_\odot$, while the
Plummer-equivalent force softening is $7.5 h^{-1}$ kpc at $z=0$.
Besides gravity and hydrodynamics, the simulation includes the
treatment of radiative cooling, the effect of a uniform
time--dependent UV background, a sub--resolution model for star
formation from a multiphase interstellar medium, as well as galactic
winds powered by SN explosions \citep{2003MNRAS.339..289S}.  At $z=0$
we extract a set of 117 clusters, whose mass, as computed from a
friends-of-friends algorithm with linking length $b=0.15$ (in units of
the mean interparticle distance) is larger than $10^{14}\msun$.
 
Due to the finite box size, the largest cluster found in the
cosmological simulation has $T_{\rm e}\approx 5$ keV.  In order to
extend our analysis to more massive and hotter systems, which are
mostly relevant for current SZ observations, we include four more
galaxy clusters having $M_{\rm vir}>10^{15} \msun$\footnote{Here
and in the following, the virial radius, $R_{\rm vir}$, is defined as
the radius of a sphere centered on the local minimum of the potential,
containing an average density, $\rho_{\rm vir}$, equal to that
predicted by the spherical collapse model. For the cosmology assumed
in our simulations it is $\rho_{\rm vir}\simeq 100 \rho_{\rm c}$, being
$\rho_{\rm c}$ the cosmic critical density. Accordingly, the virial mass,
$M_{\rm vir}$, is defined as the total mass contained within this
sphere.}  and belonging to a different set of hydro-N-body simulations
\citep{2006MNRAS.tmp..270B}. Since these objects have been
obtained by re-simulating, at high resolution, a patch of a
pre-existing cosmological simulation, they have a better mass
resolution, with $m_{\rm gas}= 1.69 \times 10^{8}
h^{-1}M_\odot$. These simulations have been performed by using the
same code with the same choice of the parameters defining
star--formation and feedback. The cosmological parameters also are the
same, except for a larger power spectrum normalization,
$\sigma_8=0.9$.
 
\begin{figure}
\centerline{
\psfig{file=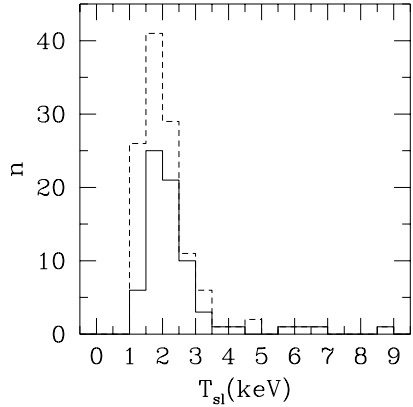,width=6cm}
}
\caption{The distribution of spectroscopic--like temperatures
for the set of simulated clusters. The dashed and the solid lines are
for the whole sample and for the subset of clusters classified as
regular (see text), respectively.}\label{fi:tmp_distr}
\end{figure}

Therefore, our total sample comprises 121 objects, spanning the 
range of spectroscopic temperatures $T_{sl}\simeq 1 - 9 $ keV, out 
of which 25 have $T_{sl} > 2.5$ keV and only four have $T_{sl} > 5$ keV. 
The corresponding temperature distribution is reported in Figure
\ref{fi:tmp_distr}. Quite apparently, our set of clusters on average samples
a lower temperature range with respect to that covered by
current SZ observations. For this reason, we will discuss in the
following the stability of our results when selecting only the high end of
the temperature distribution. Since our set of simulated clusters
covers a relatively low temperature range, we can safely ignore any
relativistic corrections to the SZ signal
\citep[e.g.,][]{1998ApJ...502....7I}.

\begin{figure}
\centerline{
\psfig{file=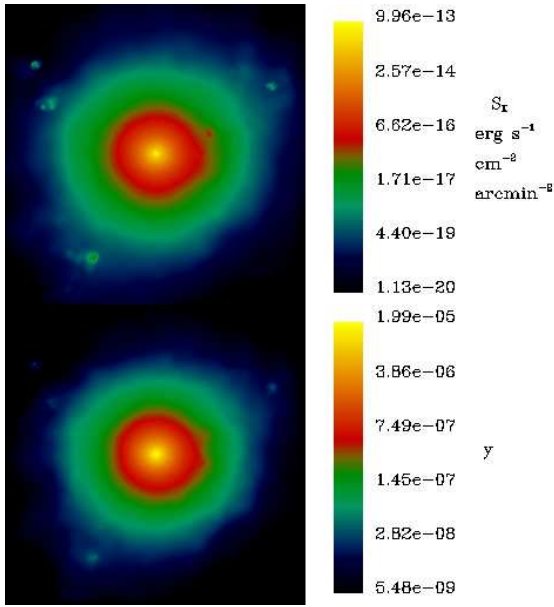,width=8cm}
}
\caption{Maps of the X-ray brightness and SZ $y$ parameter for a
regular simulated cluster having virial mass $M_{vir} \simeq 1.4 \cdot
10^{14} h^{-1}M_\odot$, $R_{500} = 0.53 h^{-1}Mps$ and $T_{sl}=2.2$
keV. The map extends out to 2 $R_{\rm vir}$, so that it covers a
physical scale of 6.05 $\hm$ for this cluster. Each map is done with a
$256\times 256$ pixelization.}\label{fi:maps1}
\end{figure}
\begin{figure}
\centerline{
\psfig{file=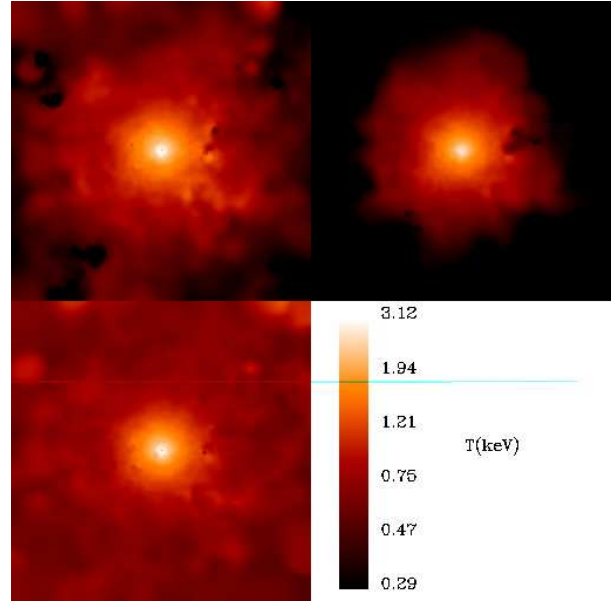,width=8cm}
}
\caption{Maps of the emission-weighted (top left), mass-weighted (top
right) and spectroscopic-like (bottom left) temperature for the same
cluster of Figure \ref{fi:maps1}.}\label{fi:maps2}
\end{figure}
Around each cluster we extract a spherical region extending out to 6
$R_{vir}$. 
Following \cite{2005MNRAS.356.1477D}, we 
create maps of the relevant quantities along three
orthogonal directions, extending out to about 2 $R_{vir}$ from the
cluster center. Each map is a regular $256\times 256$ grid.

In the Tree+SPH code, each gas particle has a smoothing length $h_i$
and the thermodynamical quantities it carries are distributed within
the sphere of radius $h_i$ according to the compact kernel: \be W(x) =
\frac{8}{\pi h_i^3} \left\{
\begin{array}{ll}
1-6x^2+6x^3 & 0 \leq x \leq \frac{1}{2}\\
2(1-x)^3    & \frac{1}{2} \leq x \leq 1\\
0           & x \geq 1
\end{array}
\right.  \ee where $x = r/h_i$ and $r$ is the distance from the
particle position.  We therefore distribute the quantity of each
particle on the grid points within the circle of radius $h_i$ centered
on the particle.  Specifically, we compute a generic quantity $q_{jk}$
on the $(j,k)$ grid point as $q_{jk}d^2_p = \int q(r) dld^2_p = \sum
q_i (m_i/\rho_i )w_i$ where $d^2_p$ is the pixel area, the sum runs
over all the particles, and $w_i \propto\int W(x) dl$ is the weight
proportional to the fraction of the particle proper volume
$m_i/\rho_i$ which contributes to the $(j,k)$ grid point. For each
particle, the weights $w_k$ are normalized to satisfy the relation
$\sum w_k = 1$ where the sum is now over the grid points within the
particle circle.  When $h_i$ is so small that the circle contains no
grid point, the particle quantity is fully assigned to the closest
grid point.  Figures \ref{fi:maps1} and \ref{fi:maps2} show an example
of the X-ray surface brightness and SZ maps and of the temperature
maps of a relaxed cluster in our simulation (with $M_{vir} \simeq 1.4
\cdot 10^{14} h^{-1}M_\odot$), which we use in the following as an
example.

\section{Results}\label{sec:results}
\begin{figure}
\centerline{
\psfig{file=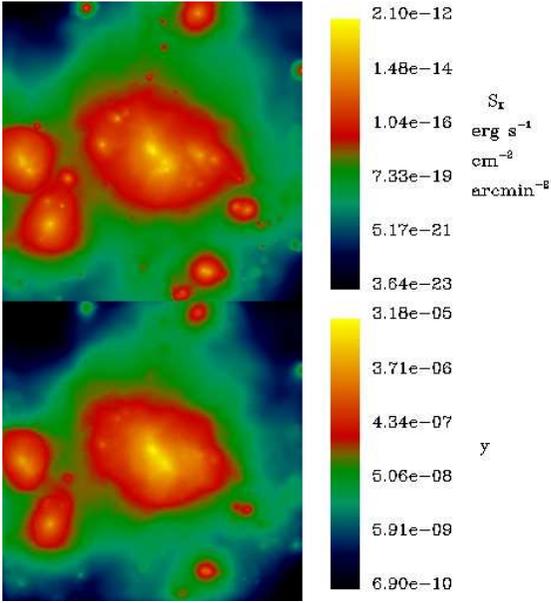,width=8cm}
}
\caption{The same as Fig.\ref{fi:maps1}, but for an
unrelaxed simulated cluster having virial mass $M_{vir} \simeq 4.1 \cdot
h^{-1}10^{14} M_\odot$. 
}\label{fi:maps3}
\end{figure}

In this section we present our results on the reliability of the
usual procedure to recover the angular--diameter distance from the
combination of the SZ and X--ray emission of clusters, by using both
the isothermal and a more general polytropic equation of state for the
ICM.  Since the procedure to determine $D_A(z)$ is known to be
particularly sensitive to the presence of cluster substructures and
irregularities, the first step of our analysis is to select a subset
of relaxed and regular clusters. Our selection criterion is based on
visual inspection of the X--ray and SZ maps, as well as on the
profiles of the X--ray surface brightness and the Comptonization
parameter. Although this is clearly not an objective criterion, it is
quite similar to the criteria used to classify real clusters. Besides
the example of a relaxed cluster shown in Figure \ref{fi:maps1}, 
in Figure \ref{fi:maps3} we also show an example of a cluster that we
classify as unrelaxed and, as such, is excluded from our analysis.
Overall, we select 71 relaxed clusters from the initial sample of 121
simulated clusters (the four hottest clusters are all included in 
this subset). 
The temperature distribution of these objects is represented
in figure \ref{fi:tmp_distr} with a solid line. We have 19 clusters 
in this subset with $T_{sl} > 2.5$ keV. We extend our statistical
sample by realizing ``observations'' of each cluster along three
orthogonal lines of sight and treating them as three independent
objects.

\subsection{Results from the isothermal model} 
Unlike X--ray observations, current observations of the SZ effect in
clusters do not allow to perform any spatially resolved analysis. For
this reason, the commonly adopted procedure is to determine the
parameters $\theta_c$ and $\beta$, which determined the $\beta$--model
density profile, from the X-ray imaging alone, along with the
normalization $S_{X0}$.  The SZ profile is then used to obtain the
central value of the Comptonization parameter, $y_0$, by using the
$\beta$-model parameters as determined from the X-ray profile. By
following this procedure, we fitted all the profiles out to $R_{500}$,
which is defined as the radius encompassing an average density of
500 times the critical cosmic density. We point out that $R_{500}$
corresponds to the typical outermost radius where X-ray observations
provide surface brightness and temperature profiles. We exclude from
the analysis the central regions of the clusters, within $1/20 R_{\rm
vir}$, which are strongly affected by gas cooling and are close to the
numerical resolution of the simulations.

\begin{figure*}
\centerline{
\psfig{file=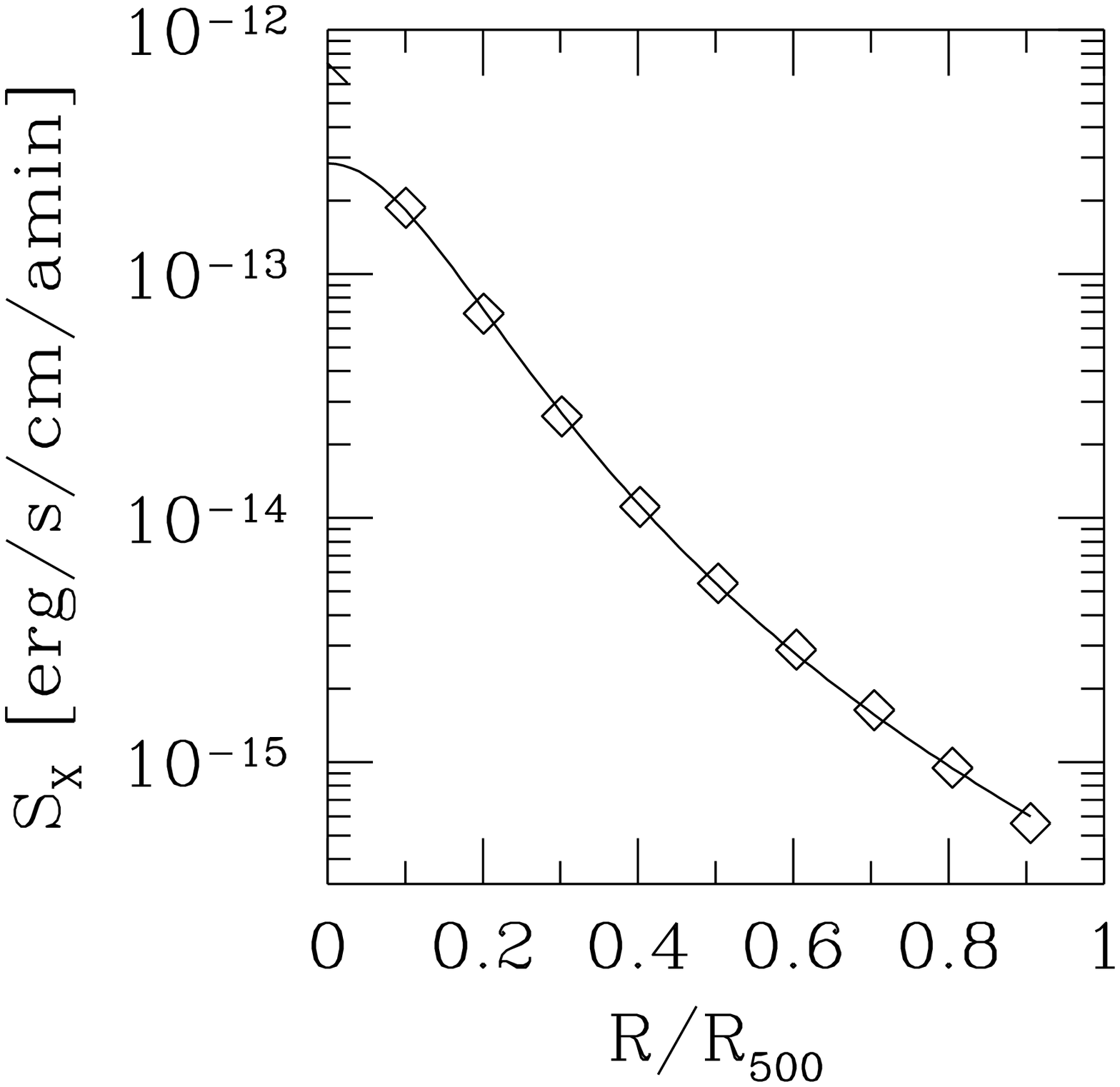,width=7cm}
\psfig{file=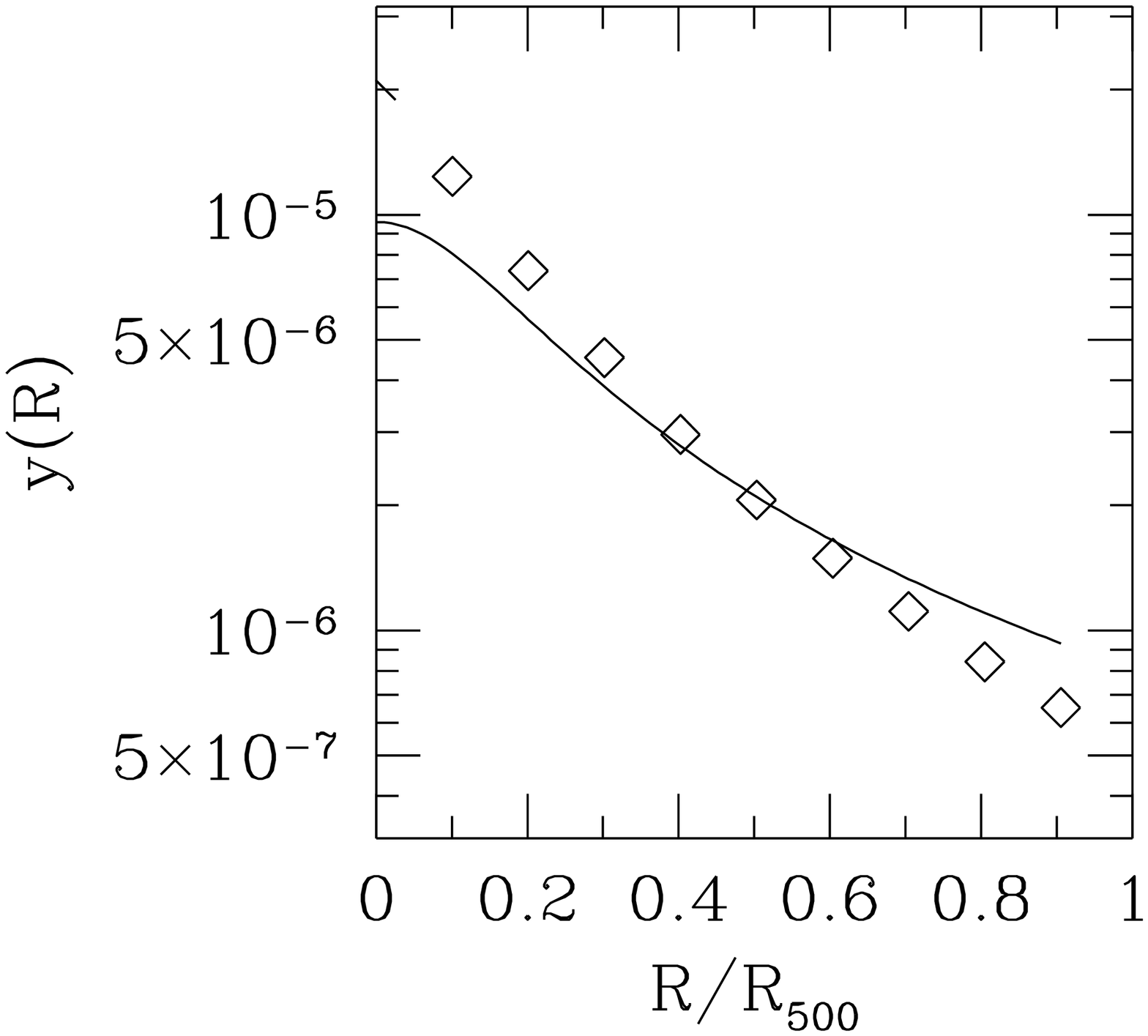,width=7cm}
}
\vspace{-.5truecm}
\caption{The projected radial profiles of $X$--ray surface brightness
  (left panel) and of Compton--$y$ parameter (right panel) for the
  cluster shown in Fig.\ref{fi:maps1}. Symbols are for the results of
  the simulation analysis while the solid curves are the predictions
  of the isothermal $\beta$--model.}\label{fi:prf1}
\end{figure*}

In Figure \ref{fi:prf1} we show the profiles of the X--ray surface
brightness and Compton--$y$ for the example cluster of
Fig.\ref{fi:maps1}, along with the best--fitting $\beta$--model for
the isothermal case. For this relaxed cluster, the $\beta$--model
provides a rather good fit to the X-ray profile. Only the central
point, which is anyway excluded from the fit, is higher than the
$\beta$--model extrapolation, as a consequence of the high--density
gas residing in the cluster cooling region. The resulting values of
the fitting parameters for this cluster are $\beta=0.835$ and
$r_c/R_{500}=0.196$. Fitting the $y(R)$ profile with
eq.(\ref{eq:ypol}), after setting $\beta=0.835$ and $\gamma=1$, leads
to an underestimate of $y_0$. In fact, the resulting Compton--$y$
profile is significantly shallower than measured (Figure
\ref{fi:prf1}).  This result follows from neglecting the presence of
negative temperature gradients.  Consequently we tend to underestimate
$D_A$, because $D_A\propto y_0^2$ [see eq.(\ref{eq:dapol})]. In this
particular case we underestimate $D_A$ by 56 per cent.

\begin{figure*}
\centerline{
\psfig{file=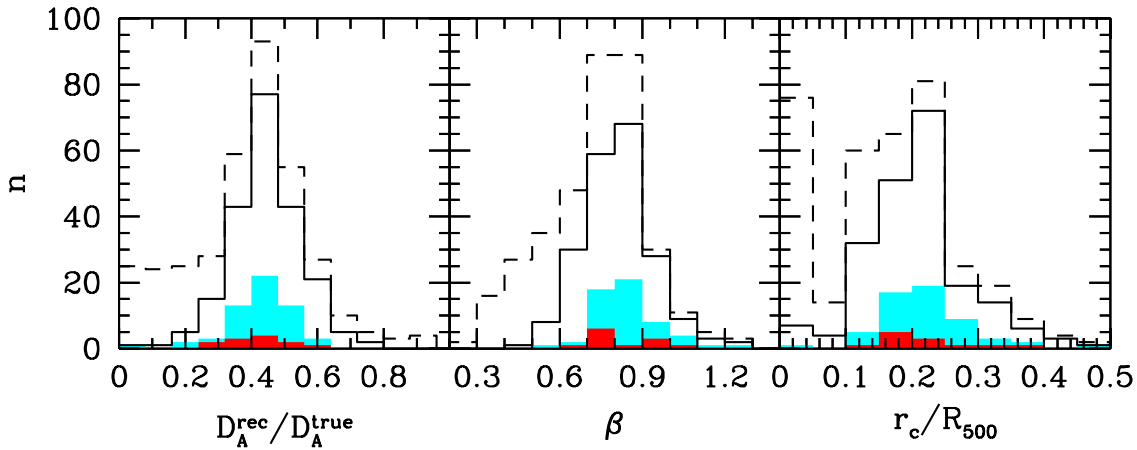,width=16cm}
}
\vspace{-8.5truecm}
\caption{The distribution of the values of $D_A^{rec}/D_A^{true}$
  (left panel), of $\beta$ (central panel) and of $r_c$ (in units of
  $R_{500}$; right panel). The dashed and the solid lines are 
  respectively for the whole sample and for the subset of clusters 
  classified as regular. Also shown with the light and dark gray areas are the
  corresponding distributions for the subset of the clusters with 
  $2.5 < T_{sl}(keV) <5$ and $T_{sl}(keV) > 5$, respectively.
  classified as regular. The distribution of $D_A^{rec}/D_A^{true}$ is
  obtained by using the electron--weighted temperatures of the
  simulated clusters in eq.(\ref{eq:dapol}).}\label{fi:res_iso}
\end{figure*}

This result for one particular cluster is confirmed by the
distribution shown in the left panel of Figure \ref{fi:res_iso} (see
also Table \ref{tab:res2}). In this figure we report the distribution
of the ratios $D_A^{\rm rec}/D_A^{\rm true}$ between the recovered and
the true values of the angular--size distance. Such results clearly
demonstrate that the angular--size distance is biased low, on average,
by more than a factor two, as a consequence of the underestimate of
$y_0$ induced by the assumption of an isothermal ICM. In order
to verify a possible temperature dependence of the $D_A$ distribution,
in the left panel of Fig.\ref{fi:res_iso} we also show the results for
the clusters with $T_{sl}$ in the range 2.5--5 keV and for those
hotter than 5 keV. While the latter are too few to allow any
meaningful conclusion, the clusters at intermediate temperature have a
distribution which is statistically consistent with that of the whole
sample. This indicates the absence of any obvious trend of our results
with the cluster size.  The results reported in this figure have
been obtained by using the electron--weighted temperature estimate for
the simulated clusters. If we had used the emission--weighted
temperature, we would have obtained an even stronger bias (see
eq.[\ref{eq:dapol}]), because this temperature generally overestimates
the electron temperature.

Including dynamically disturbed clusters does
not significantly affect the average value of the recovered
$D_A$. However, the resulting distribution is clearly asymmetric and
presents a large tail towards low $D_A$ values.  In fact,
eqs. (\ref{eq:y}) and (\ref{eq:sx}) show that that $D_A \propto
\langle n_e\rangle^2/\langle n_e^2\rangle$. Therefore, the presence of
clumps in the gas distribution produces an underestimate of $D_A$ by
this factor with respect to a completely smooth gas distribution.
By looking at the distributions of the $\beta$ and $r_c$ (central and
left panels of Fig.\ref{fi:res_iso}), unrelaxed structures tend to
have rather flat gas density profiles. Fitting them with a
$\beta$--model forces the slope to be very small, with a preference
for the core radius to be consistent with zero. For instance, the
cluster shown in Fig.\ref{fi:maps3} requires $\beta=0.52$ and
$r_c/R_{500}=0.03$, while its estimate of the angular--size distance
gives $D_A^{\rm rec}/D_A^{\rm true}=0.17$.

As a word of caution in interpreting such results, we emphasize that
this bias in the $D_A$ estimate, due to the isothermal gas assumption, is
likely to represent an
overestimate of the actual effect in real cluster observations for at
least two reasons. First, radiative simulations of clusters are
known to produce temperature gradients that, in the central regions,
are steeper than observed (see the discussion in \S 2.1). As a
consequence, simulated clusters exaggerate the departure from
isothermality. Second, the $\beta$--model fitting to the Compton--$y$
profile has been performed by assigning equal weight to all radial
bins, with the more external regions bringing down the overall
normalization of the model profile. In a realistic observational
setup, the signal from central cluster regions should have a
relatively larger weight, thus reducing the bias in the recovered 
$y_0$. Addressing appropriately this issue would require implementing 
detailed mock SZ observations of our simulated clusters, a task that
we defer to a future analyses. Even keeping in mind these warnings, it
is clear that deviations from isothermality must be taken into
account for a precise calibration of the cosmic distance scale from
the combination of X--ray and SZ observations of galaxy clusters
\citep[e.g.,][]{2004ApJ...615...63U}.

\begin{table}
\begin{center}
\begin{tabular}{lcc}
\hline
& All  & Regular  \\
& clusters & clusters  \\
\hline
\hline
Isothermal  &$  0.41  \pm  0.27  $&$  0.44  \pm  0.11  $\\        
  &$  0.42  _{-  0.22  }^{+  0.14  }$&$  0.45  _{-  0.10  }^{+  0.10  }$\\
  \hline
Polytropic & & \\ 
$T_{ew}$   &$  0.76  \pm  0.74  $&$  0.80  \pm  0.15  $\\        
  &$  0.78  _{-  0.37  }^{+  0.19  }$&$  0.81  _{-  0.17  }^{+  0.14  }$\\
$T_{e}$   &$  0.95  \pm  0.49  $&$  1.04  \pm  0.22  $\\        
  &$  0.98  _{-  0.47  }^{+  0.31  }$&$  1.05  _{-  0.26  }^{+  0.21  }$\\
$T_{sl}$   &$  0.99  \pm  2.57  $&$  0.97  \pm  0.18  $\\        
  &$  0.92  _{-  0.46  }^{+  0.25  }$&$  0.98  _{-  0.20  }^{+  0.18  }$\\
\hline
\end{tabular}
\end{center}
\caption{The values of the accuracy in recovering the angular--diameter
  distance, $D_A^{\rm rec}/D_A^{\rm true}$, using both the isothermal
  and the polytropic model, and using the emission--weighted, the
  electron and the spectroscopic--like definitions of temperature. For
  each of them, the first line reports the mean and standard
  deviation, the second the median and the limiting values
  encompassing 68\% of the data.}\label{tab:res2}
\end{table}

\subsection{Results from the polytropic fit}
In the case of a more general polytropic equation of state, the
parameters $\beta$ and $\gamma$ are calculated by requiring the model
to reproduce at the same time both the X-ray surface brightness
and the temperature profiles. After obtaining the core radius $r_c$
and the normalization $S_{X0}$ from the X-ray profile, and $T_0$ from
the temperature profile, we combine the two exponents in
eqs. (\ref{eq:sxpol}) and (\ref{eq:tpol}) to derive both $\beta$ and
$\gamma$, with $y_0$ finally obtained from the SZ profile.

\begin{figure*}
\centerline{
\psfig{file=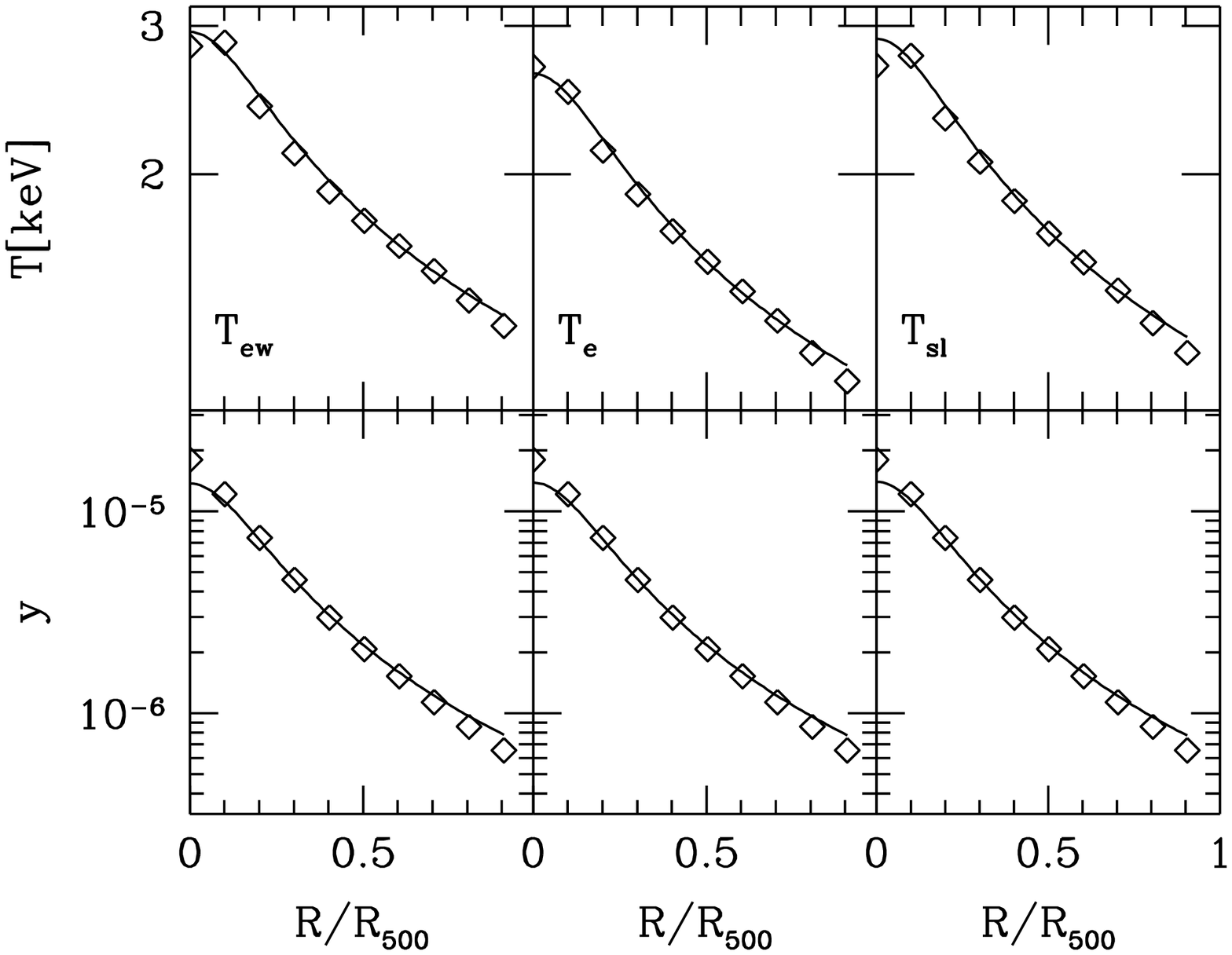,width=11.cm}
}
\vspace{-2.5truecm}
\caption{Profiles of temperature (upper panels) and of Compton--$y$
  parameter (lower panels) for the cluster of
  Fig.\ref{fi:maps1}. Left, central and right panels corresponds to
  using emission--weighted, electron and spectroscopic--like
  temperature, respectively. Open symbols are for the profiles from
  the simulation analysis, while the curves are the best--fitting
  polytropic $\beta$--model.  }\label{fi:tmp4}
\end{figure*}

\begin{figure*}
\centerline{
\psfig{file=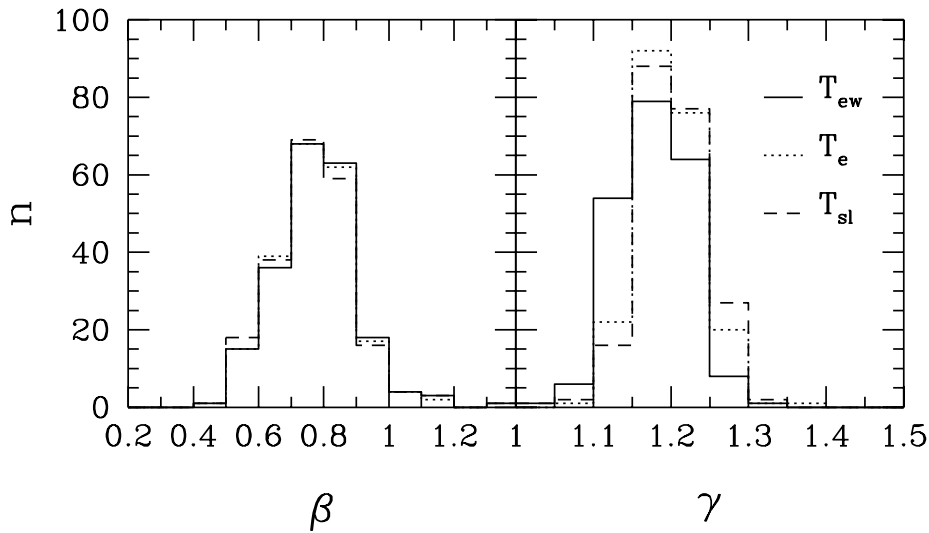,width=13cm}
}
\vspace{-5.0truecm}
\caption{Distributions of the values of 
$\beta$ (left panel) and $\gamma$ (right panel), using
emission--weighted (solid line), electron (dotted line) and
spectroscopic--like (dashed lines) temperatures, respectively, as
obtained for the whole sample of 121 clusters.}\label{fi:betagamma}
\end{figure*}

In Figure \ref{fi:tmp4} we show the temperature and Compton--$y$
profiles for our example cluster, along with the best--fitting
predictions of the polytropic $\beta$--model, for the three different
definitions of temperature. The polytropic equation of state provides
a reasonable approximation to all temperature profiles and, unlike 
the isothermal case, allows us to correctly predict also the
Compton--$y$ profile. The corresponding distributions of $\beta$ and
$\gamma$ are shown in Figure \ref{fi:betagamma} (we do not report the
distribution of $r_c$, since it is, by definition, identical to that of
the isothermal model). For both quantities, the effect of using
different definitions of temperature is rather small. As expected,
using a polytropic temperature profile implies only a modest
decrease of the $\beta$ values, because of the weak temperature
dependence of the cooling function. All the three distributions of $\gamma$ 
have an average value $\simeq 1.2$, similar to
observational estimates \citep[e.g.][]{2002ApJ...567..163D}. Moreover, the
scatter in this distribution is so small to make the isothermal ICM an
extremely unlikely event.

\begin{figure*}
\centerline{
\psfig{file=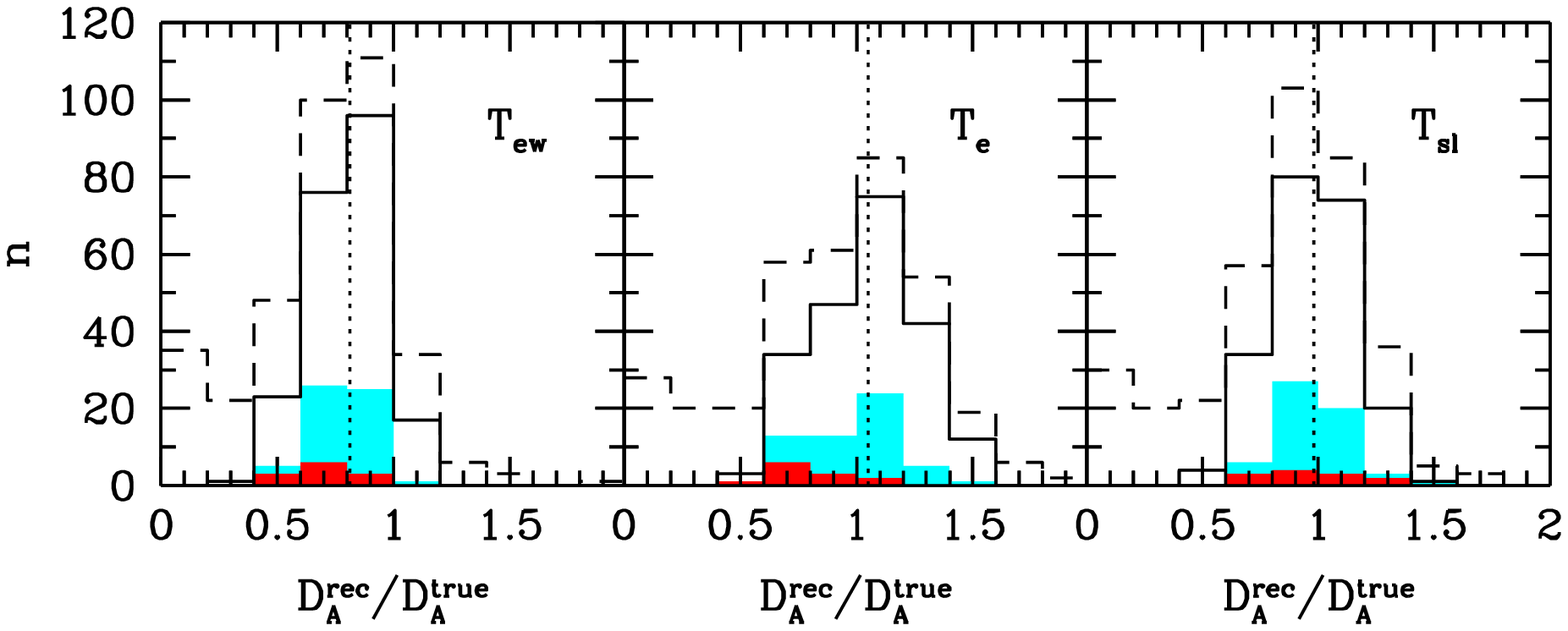,width=15cm}
}
\vspace{-8.truecm}
\caption{Distributions of the accuracy in recovering the correct value
  of the angular--diameter distance, $D_A^{\rm rec}/D_A^{\rm true}$, using
  the polytropic $\beta$--model for the whole sample (dashed line)
  and for the subset of regular clusters (solid), using
  emission--weighted (left panel), electron (central panel) and
  spectroscopic--like (right panel) temperature. Also shown with the 
  light and dark gray areas are the corresponding distributions for the 
  subset of the clusters with $2.5 < T_{sl}(keV) <5$ and 
  $T_{sl}(keV) > 5$, respectively. The vertical dotted line in
  each histogram represents the mean value of the distribution for the
  sample of regular clusters.}
\label{fi:da}
\end{figure*}

The results obtained for $D_A$ are shown in Figure \ref{fi:da}, and
also reported in Table \ref{tab:res2}, using emission--weighted, electron
and spectroscopic--like temperatures. Quite 
interestingly, the improved quality of the fit to the profile of the Compton--$y$
parameter now makes the distribution peak at a value much closer to
the correct $D_A$, independently of whether we use the whole sample or
the subsample of relaxed clusters.

The angular--diameter distance is correctly recovered when using either the
electron or the spectroscopic--like temperature with deviations which
are always $\mincir 5$ per cent, on average. 
This is a rather encouraging result, since it indicates that any bias,
induced by using the temperature as measured from X--ray observations,
is in fact rather small. Using instead the emission--weighted
temperature turns into a systematic underestimate of $D_A$ by about 20
per cent, as a consequence of the fact that it is systematically
higher than the electron temperature. For all the definitions of
temperature we find a significant intrinsic scatter, of about 20 per
cent on average, in spite of our selection of regular objects.

The fact that the scatter is stable against the definition of
temperature implies that it is almost insensitive to the thermal
structure of the ICM and, therefore, to the details of the ICM
physics. This scatter instead quantifies the effect of departure from 
spherical symmetry of the ICM spatial distribution. 
In fact, the above scatter increases to about 50 per cent, if no
preselection of regular clusters is implemented (Table \ref{tab:res2}). Quite remarkably, the
intrinsic scatter calibrated with our simulations is rather close to
the 17 per cent value, reported by \cite{1998ApJ...501....1H}, for the
uncertainty induced by the intrinsic cluster ellipticity. 

Similarly to the case of the isothermal fit, we note from
Fig. \ref{fi:da} that the low--$D_A$ tails of the distributions are
contaminated by irregular clusters, for which $D_A$ is very badly
recovered. For instance, for the irregular cluster shown in
Fig. \ref{fi:maps3} we find $D_A^{\rm rec}/D_A^{\rm true}=0.30$ when
using the electron temperature. Similarly to the case of
the isothermal fit, also in this case the distribution of the hot
clusters is consistent with that of the regular cluster subset.

\begin{figure*}
\centerline{
\psfig{file=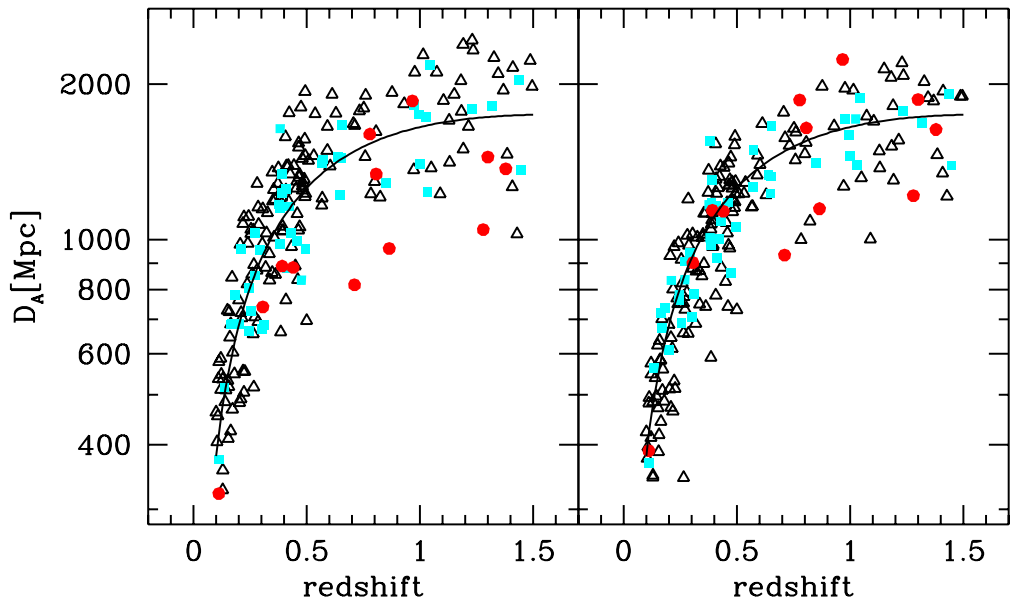,width=15cm}
}
\vspace{-6.truecm}
\caption{Estimated $D_A$ vs. $z$ for the regular cluster sample: 2/3 of this 
sample was uniformly distributed in the redshift range $0.1<z<0.5$ and 1/3 
in the range $0.5<z<1.5$. On the left (right) panel is shown the angular--size
distance obtained using the electron (spectroscopic-like) temperature and the
polytropic gas model. The symbols indicate clusters spectroscopic temperature: 
$T_{sl}(keV) < 2.5$ (triangles), $2.5 < T_{sl}(keV) < 5$ (squares) and
$T_{sl}(keV) > 5$ (circles).
The solid line shows the $D_A$--$z$ relation for the $\Lambda$CDM
cosmology assumed in the simulations.
}
\label{fi:da-z}
\end{figure*}

\subsection{Implications for cosmological parameters}
The precision in the recovery of the angular--size distance when using
the polytropic model indicates that this method is potentially
accurate to estimate cosmological parameters. In order to test this we
create a simple mock catalog of clusters, which is obtained by
distributing 2/3 of our regular clusters uniformly in redshift in the
range $0.1<z<0.5$, while the remaining 1/3 is distributed uniformly in
the range $0.5<z<1.5$.  Recall that each simulated cluster is observed
along three orthogonal lines of sight and the redshift of each
projection is chosen randomly.  Figure \ref{fi:da-z} shows the
resulting distribution of clusters in the $D_A$--$z$ plane. We
remind here that our simulated clusters have been identified at
$z=0$. Therefore, our procedure to distribute them at $z>0$ neglects
the effect of their possible morphological evolution. We have been
forced to this choice by the small volume of our simulation box,
which implies the rapid disappearance of reasonably massive
clusters inside the high--redshift simulation box.

For the estimate of the Hubble constant, $H_0$, we limit the analysis
to the 66 clusters at $z<0.3$.  Including high redshift objects would
make the recovery of $H_0$ progressively more dependent on the
knowledge of the underlying cosmology. When using the electron
temperature, the distribution of the $H_0$ values has mean $H_0=70\pm
2\hub$; when using the spectroscopic temperature this mean is
$H_0=75\pm 2\hub$. In both cases the uncertainties are the $1-\sigma$
standard deviations. These values are obtained by assuming the correct
cosmology. When assuming the Einstein--de-Sitter model, we find $H_0$
biased low by 8 per cent.

As for the estimate of the matter density parameter $\Omega_m$, we fix the
value of $H_0$ to its true value and assume flat geometry. In this
case, we use the 73 clusters lying at $z>0.5$. Estimating $\Omega_m$
as the average of the values yielded by each cluster would provide
unreliable results; in fact, inaccurate values of $D_A$ can imply
negative values of $\Omega_m$, which are clearly
unphysical. Therefore, we compute the best--fitting value of
$\Omega_m$ with a $\chi^2$--minimization procedure. To associate the
uncertainty to the estimated $\Omega_m$, we resort to a bootstrap
resampling procedure (e.g. \S 15.6 of
\citealt{1992nrfa.book.....P}). Each bootstrap sample is constructed
by randomly selecting, with repetition, the objects from the original
sample. Each time that a cluster is selected, its $D_A$ is perturbed
with a Gaussian random shift with variance 20\%, independently of
redshift, to account for the ``observational'' uncertainties. The
application of this procedure, when using the electron temperature,
gives $\Omega_m=0.29\pm 0.05$; we obtain $\Omega_m=0.36\pm 0.06$, when
using the spectroscopic--like temperature. The uncertainties are the
$1-\sigma$ standard deviations computed with 100 bootstrap
resamplings.  The two temperature definitions provide two $\Omega_m$'s
whose difference is consistent with the difference in the median
values of $D_A$.  Moreover, and reassuringly, in both cases the
central values are consistent with the true value of $\Omega_m$.

The small size of the errobars of the estimated $\Omega_m$'s
should be clearly taken with caution for at least two reasons. First
of all, we have assumed errors in $D_A$ to be 20 per cent,
independently of redshift. High--quality SZ and X--ray observations will
eventually allow to bring statistical errors down
to this level in the near future. Of course, systematic errors in SZ observations, 
associated for instance to
point--source contamination and CMB signal removal, 
are different in nature and more difficult to eliminate.

\section{Conclusions}
In this paper we have applied the method to calibrate the cosmic
distance scale from the combination of X--ray and Sunyaev--Zeldovich
(SZ) observations to an extended set of hydrodynamical
simulations of galaxy clusters. The simulations have been performed with the
{\small GADGET2} code, for a flat $\Lambda$CDM model with
$\Omega_m=0.3$, $h=0.7$ and $\sigma_8=0.8$, and include the effect of
cooling, star formation and supernova feedback. The aim of our
analysis was to understand the possible biases introduced by the
assumptions of isothermal gas and the X--ray temperature
as a close proxy to the electron temperature, as usually done in
the analysis of real clusters. Furthermore, the application of this
method to a large set of simulated clusters allows us to quantify the
intrinsic scatter associated with a cluster-by-cluster variation of
their shapes.

Our main results can be summarized as follows.

\begin{description}
\item[(a)] Neglecting the temperature gradients in the application of
  the $\beta$--model produces a significant underestimate of the
  central value of the Comptonization parameter, $y_0$. In turn, this
  introduces a severe bias in the estimate of the angular--size
  distance, $D_A$. 
\item[(b)] Accounting for the presence of the temperature gradients
  with a polytropic $\beta$--model substantially reduces this bias to
  a few per cent level. While this result holds when using either the
  electron or the spectroscopic--like temperature, using the
  emission--weighted temperature gives a $\sim 20$ per
  cent underestimate of $D_A$.
\item[(c)] Cluster-by-cluster variations of the asphericity and of the
  degree of gas clumpiness cause an intrinsic dispersion of about
  $\sim 20$ per cent in the estimates of $D_A$. This dispersion
  significantly increases in case unrelaxed clusters are included in
  the analysis.
\item[(d)] The set of simulated clusters is used to generate a mock
  sample of clusters out to redshift $z=1.5$. By assuming a 20 per
  cent precision in the estimate of $D_A$ for each cluster, we find
  that the correct value of $H_0$ is recovered with a statistical
  error of $2\vel$ at 1$\sigma$. Furthermore, assuming a prior for the
  Hubble constant and flat geometry, we find that also the matter
  density parameter can be estimated in an unbiased way with a
  statistical error of $\Delta \Omega_m= 0.05$.
\end{description}

It is worth reminding here that our results are based on the analysis
of simulated X--ray and SZ maps, which are ideal in a number of
ways. First of all, they have been generated by projecting the signal
contributed by the gas out to about six virial radii. A more rigorous
approach would require projecting over the cosmological light cone, to
properly account for the fore/background contamination. While
projection effects ought to be marginal for the X--ray maps, they may
substantially affect the SZ signal
\citep[e.g.,][]{2002ApJ...579...16W,2005astro.ph.11357D}. Furthermore, our noiseless maps
need to be properly convolved with the ``response function'' of both
X--ray and SZ telescopes under realistic observing
conditions. Neglecting the observational noise clearly leads to an
underestimate of the uncertainties in the determination of the
parameters defining gas density and temperature profiles. Accounting
for such effects would definitely require passing our ideal maps
through suitable tools to simulate X--ray
\citep[e.g.,][]{2004MNRAS.351..505G} and SZ observations
\citep[e.g.,][]{2001MNRAS.328..783K,2005MNRAS.359..261P}. Finally, the
effect of neglecting the departure from isothermality depends on the
physical description of the ICM provided by the simulations. Since
simulated clusters have central temperature gradients, which are
steeper than the observed ones, the above effect is probably
overestimated. This demonstrates that a proper use of hydrodynamical
simulations to calibrate galaxy clusters as standard rod also requires
a correct description of the physical properties of the intra--cluster
gas.

When this paper was ready for submission it came to our attention a
paper by \citet{2005astro.ph.10745H}, which reports on an analysis
similar to ours. However, their results indicate that assuming an
isothermal gas tends to overestimate rather underestimate $D_A$, as we
find. Clearly, a detailed comparison of their approach with ours would
be necessary to understand this discrepancy.  We also note that these
authors propose a novel method to estimate $D_A$ which relies on
spatially resolved SZ observations. On the contrary, our suggestion of
considering a polytropic gas model can also be applied when the
spatial information of the SZ data is poor.

\section*{Acknowledgments.}
The simulations have been performed using the IBM--SP4 machine at the
``Consorzio Interuniversitario del Nord-Est per il Calcolo
Elettronico'' (CINECA, Bologna), with CPU time assigned thanks to the
INAF--CINECA grant, and the IBM--SP4 machine at the ``Rechenzentrum
der Max-Planck-Gesellschaft'' at the ``Max-Planck-Institut f\"ur
Plasmaphysik'' with CPU time assigned to the ``Max-Planck-Institut
f\"ur Astrophysik''. We wish to thank an anonymous referee for
detailed comments, which helped improving the presentation of the
results.  We would like to thank Giuseppe Murante for his help in the
initial phase of this project, and Stefano Ettori for enlightening
discussions. This work has been partially supported by the INFN--PD51
grant and by MIUR. Part of this work served as the master degree 
thesis of S. A. at the ``Universit\`a degli Studi di Torino''.

\bibliographystyle{mn2e}
\bibliography{master}

\end{document}